\documentclass[aps,prd,twocolumn,superscriptaddress,nofootinbib,10pt]{revtex4-2}

\usepackage{amsmath,amssymb,bm}
\usepackage{graphicx}
\usepackage{booktabs}
\usepackage[colorlinks=true,linkcolor=blue,citecolor=blue,urlcolor=blue]{hyperref}

\newcommand{\MGUT}{M_{\rm GUT}}
\newcommand{\MPl}{M_{\rm Pl}}

\begin{document}

\title{Testing the Unitarity of Three-Flavor Neutrino Mixing above the GUT Scale}

\author{Bipin Singh Koranga}
\email{bskoranga@kmc.du.ac.in}
\affiliation{Department of Physics, Kirori Mal College, University of Delhi, Delhi--110007, India}

\author{Samrat Vashisht}
\affiliation{Department of Physics, Kirori Mal College, University of Delhi, Delhi--110007, India}

\date{\today}

\begin{abstract}
We examine how physics between the grand-unification scale and the Planck scale can
affect the unitarity of the neutrino mixing matrix in the standard three-flavor
framework. Two distinct effects are separated. First, very heavy singlet states at or
above the grand-unification scale make the $3\times3$ PMNS matrix non-unitary, but only
at a level set by the ratio of the light-neutrino mass to the heavy mass; we find this
to be about $10^{-27}$, roughly twenty-five orders of magnitude below the present
oscillation bound. Second, a flavor-blind dimension-five (Weinberg) operator suppressed
by a high scale $\Lambda$ shifts the mixing angles while leaving the $3\times3$ matrix
exactly unitary. Using the NuFIT~6.1 normal-ordering best fit, including the Dirac CP
phase, and diagonalizing the corrected complex-symmetric mass matrix numerically, we
find shifts of $\theta_{12}$ of at most about $0.12^\circ$ at the Planck scale for a
quasi-degenerate spectrum and about $0.011^\circ$ for the normal ordering, growing as
$1/\Lambda$ to several degrees near the grand-unification scale ($6.5^\circ$ for the
normal ordering), where perturbation theory fails for the quasi-degenerate case. Because
the operator is gauge invariant in the three-flavor basis and no sterile sector is
present, the shifts induce no change in any non-unitarity parameter; instead the three
angle shifts are tied together by a single parameter. A unitarity test based on
conservation of total probability among the active flavors is therefore not a probe of
the scale above grand unification; the relevant information resides in the unitary
shifts of the mixing angles.
\end{abstract}

\maketitle

\section{Introduction}
\label{sec:intro}

Neutrino oscillation data from solar, atmospheric, reactor, and accelerator experiments
are described to good accuracy by a unitary $3\times3$
Pontecorvo--Maki--Nakagawa--Sakata (PMNS) matrix~\cite{pontecorvo1968,maki1962},
parametrized by three mixing angles, one Dirac CP phase (and, for Majorana neutrinos,
two further phases that do not affect oscillations), together with two independent
mass-squared differences; current global fits and parameter ranges are given in
Refs.~\cite{pdg2024,esteban2020,esteban2024,nufit61}.

Unitarity of this matrix is an assumption, not a theorem of the data. It fails whenever
additional states mix with the active neutrinos but are not kinematically accessible,
for example because they are too heavy to be produced or because their oscillation
phases average out~\cite{antusch2006,escrihuela2015,blennow2017}. Light sterile states,
by contrast, are accommodated by enlarging the mixing matrix, which stays
unitary~\cite{giunti2000,nautiyal2021a,nautiyal2021b,koranga2026}. Non-unitarity of the
active block can be tested through zero-distance effects and through the sum rule
\begin{equation}
  \sum_{\beta=e,\mu,\tau} P_{\alpha\beta} = 1 - \epsilon_\alpha,
  \label{eq:sumrule}
\end{equation}
where $\epsilon_\alpha$ may in general depend on the neutrino energy and baseline, as
discussed in Refs.~\cite{giunti2019,unitarityframework}. Existing bounds, for example
$\epsilon_\mu \le 0.064$ and $\epsilon_\tau \le 0.44$ at $3\sigma$~\cite{unitarityframework},
are fairly weak and constrain new states that mix at the percent level, which in a
seesaw-like picture means masses far below the grand-unification scale
$\MGUT \sim 2\times10^{16}$~GeV.

Grand unification and quantum gravity are both expected to leave imprints on the lepton
sector, but those imprints are conventionally associated with extremely heavy states, as
in the seesaw mechanism~\cite{minkowski1977,yanagida1979,gellmann1979,mohapatra1980}, or
with high-dimension operators, including those induced by Planck-scale
physics~\cite{barbieri1980,akhmedov1992}. The question addressed here is: what can a
unitarity test of the three-flavor mixing matrix say about the regime above the GUT
scale? To answer it we separate two mechanisms that are often conflated, namely
non-unitarity from heavy states and unitary rotations from higher-dimension operators,
and we quantify each of them. Working with three flavors has two simplifying
consequences: the Weinberg operator acts on exactly the fields that appear in the mixing
matrix, so the perturbation is gauge invariant without further qualification, and the
only source of non-unitarity is the decoupled heavy sector.

Section~\ref{sec:framework} sets up the framework, Sec.~\ref{sec:numerics} presents the
numerical diagonalization, Sec.~\ref{sec:discussion} discusses the results and
limitations, and Sec.~\ref{sec:conclusion} concludes.

\section{Framework}
\label{sec:framework}

\subsection{Three-flavor mixing}

We write the flavor states as superpositions of mass eigenstates,
$\nu_\alpha = \sum_{i=1}^{3} U_{\alpha i}\,\nu_i$ with $\alpha = e,\mu,\tau$, and use the
standard parametrization
\begin{equation}
  U = R_{23}(\theta_{23})\, R_{13}(\theta_{13},\delta)\, R_{12}(\theta_{12}),
  \label{eq:param}
\end{equation}
where $R_{ij}$ is a rotation in the $ij$ plane. In the numerical illustration of
Sec.~\ref{sec:numerics} the Dirac CP phase is set to its best-fit value, so $U$ is a
complex unitary matrix; the Majorana phases are set to zero. By construction
$U^\dagger U = \mathbb{1}$ and $\sum_{i=1}^{3}|U_{\alpha i}|^2 = 1$ for every $\alpha$.
To quantify departures from unitarity we define the row deficit
\begin{equation}
  \epsilon_\alpha = 1 - \sum_{i=1}^{3} |U_{\alpha i}|^2,
  \label{eq:deficit}
\end{equation}
which is exactly zero for Eq.~\eqref{eq:param} and nonzero only if the $3\times3$ matrix
is the light block of a larger unitary matrix. In a $(3+1)$ scheme this deficit equals
the sterile fraction $|U_{\alpha4}|^2$; here, with no light sterile state, any nonzero
$\epsilon_\alpha$ must come from states too heavy to take part in oscillations.

\subsection{Heavy states above $\MGUT$}

Consider a heavy gauge-singlet state $N$ of mass $M$ far above any accessible energy,
with Dirac mass $(m_D)_\alpha$ to the active neutrinos. In the basis $(\nu_\alpha, N)$
the mass matrix has the block form
\begin{equation}
  \begin{pmatrix} 0 & m_D \\ m_D^{T} & M \end{pmatrix}.
  \label{eq:massblock}
\end{equation}
For $M \gg m_D$, integrating out the heavy state gives the usual seesaw
relation~\cite{minkowski1977,yanagida1979,gellmann1979,mohapatra1980}
$m_\nu \simeq m_D^2/M$ for the light mass, and the decoupling rotation leaves an
active--heavy mixing $|U_{\alpha H}| \simeq m_D/M$. The light block of the full unitary
matrix is $U_{\rm light} = (\mathbb{1}-\eta)\,U$ with
$\eta_{\alpha\beta} \simeq \tfrac12 \sum_H U_{\alpha H} U^*_{\beta H}$, so that, in the
standard parametrization of non-unitary mixing~\cite{antusch2006,escrihuela2015}, the
row deficit is
\begin{equation}
  \epsilon_\alpha \simeq \frac{|(m_D)_\alpha|^2}{M^2}
  = \frac{m_\nu}{M}\,\frac{|(m_D)_\alpha|^2}{m_D^2} \lesssim \frac{m_\nu}{M},
  \label{eq:epsheavy}
\end{equation}
where $m_D^2 \equiv \sum_\alpha |(m_D)_\alpha|^2$ and $m_\nu = m_D^2/M$. Because $m_\nu$
is sub-eV, the non-unitarity falls inversely with $M$. For $m_\nu = 0.05$~eV and
$M = \MGUT$, Eq.~\eqref{eq:epsheavy} gives $\epsilon \sim 2.5\times10^{-27}$
($0.05~\text{eV} = 5\times10^{-11}$~GeV, divided by $2\times10^{16}$~GeV). Since the
heavy state cannot be produced, its only footprint in oscillations is the constant,
energy-independent deficit $\sum_\beta P_{\alpha\beta} = 1-\epsilon_\alpha$, the
zero-distance effect. It cannot be distinguished from a normalization error except by
comparing near and far detectors or by combining channels, which is why the bounds of
Ref.~\cite{unitarityframework} require a global treatment.

Equation~\eqref{eq:epsheavy} assumes a single-state, unsuppressed seesaw in which the
light mass arises solely from $m_D^2/M$. With several heavy states, or with a symmetry
that makes the light mass much smaller than $m_D^2/M$ (as in inverse- or linear-seesaw
constructions, for which the non-unitarity bounds of Refs.~\cite{antusch2006,blennow2017}
apply directly), $m_D/M$ can exceed $\sqrt{m_\nu/M}$ and the estimate no longer applies;
we return to this in Sec.~\ref{sec:discussion}.

\subsection{Planck-type corrections in the three-flavor basis}

The lowest-dimension operator that can be added to the Standard Model above the
electroweak scale is the dimension-five Weinberg operator~\cite{weinberg1979}, which is
also the leading effect expected from Planck-scale physics~\cite{barbieri1980,akhmedov1992},
\begin{equation}
  \mathcal{L}_5 = \frac{c_{\alpha\beta}}{\Lambda}
  \left(\overline{L^c_\alpha}\,\tilde H^{*}\right)\left(\tilde H^{\dagger} L_\beta\right)
  + \text{h.c.},
  \label{eq:weinberg}
\end{equation}
where $L_\alpha$ is the lepton doublet of flavor $\alpha$,
$L^c_\alpha = C\bar L_\alpha^{T}$ its charge conjugate, $\tilde H = i\sigma_2 H^*$, and
$c_{\alpha\beta}$ is a symmetric matrix of dimensionless coefficients. After electroweak
symmetry breaking, $\langle H\rangle = v$, it contributes a Majorana mass matrix
$c_{\alpha\beta}v^2/\Lambda$. In the three-flavor framework every field in
Eq.~\eqref{eq:weinberg} is an active lepton doublet, so the operator is invariant under
$SU(2)_L\times U(1)_Y$ with no special treatment: there is no sterile row or column that
would have to be excluded.

If gravity is flavor blind, the coefficients are universal, $c_{\alpha\beta}=c$, and the
contribution is proportional to the $3\times3$ all-ones matrix $\mathbf{J}$. With the
convention in which the unperturbed mass matrix is diagonalized as
$U_0^{T}\mathcal{M}_0U_0=\mathrm{diag}(m_1,m_2,m_3)$, the corrected mass matrix in the
flavor basis is
\begin{equation}
  \mathcal{M} = U_0^{*}\,\mathrm{diag}(m_1,m_2,m_3)\,U_0^{\dagger} + \mu\,\mathbf{J},
  \qquad \mu = \frac{c\,v^2}{\Lambda},
  \label{eq:corrmass}
\end{equation}
where $U_0$ is the unperturbed matrix of Eq.~\eqref{eq:param} and $v = 174$~GeV. (For real
$U_0$ this coincides with $U_0\,\mathrm{diag}(m_i)\,U_0^{T}$; with a nonzero Dirac phase
the two forms are related by complex conjugation of $U_0$.)
Numerically $\mu \simeq 2.5\times10^{-6}$~eV at $\Lambda = \MPl = 1.22\times10^{19}$~GeV
and $\mu \simeq 1.5\times10^{-3}$~eV at $\Lambda = 2\times10^{16}$~GeV.

For $\delta\neq0$ the matrix $\mathcal{M}$ is complex symmetric, and it is brought to
diagonal form with non-negative entries by a Takagi factorization,
$U'^{T}\mathcal{M}\,U'=\mathrm{diag}(m_1',m_2',m_3')$, with $U'$ unitary. The row sums
\begin{equation}
  \sum_{i=1}^{3} |U'_{\alpha i}|^2 = 1
  \label{eq:rowsum}
\end{equation}
are therefore preserved identically. There is no fourth state into which weight can be
redistributed, so $\epsilon'_\alpha = \epsilon_\alpha = 0$: the effect rearranges the
mixing without opening the system.

To first order in $\mu$ write $U'=U_0(\mathbb{1}+T)$ with $T$ anti-Hermitian. Defining
\begin{equation}
  B_{ij}\equiv(U_0^{T}\,\mu\mathbf{J}\,U_0)_{ij}=\mu\,N_iN_j,
  \qquad
  N_i \equiv \sum_{\alpha=e,\mu,\tau} U_{0,\alpha i},
  \label{eq:Ni}
\end{equation}
the Takagi condition gives, for $i\neq j$,
\begin{equation}
  T_{ij} = -\frac{\mathrm{Re}\,B_{ij}}{m_i-m_j}
  - i\,\frac{\mathrm{Im}\,B_{ij}}{m_i+m_j},
  \qquad
  T_{ii} = -\,i\,\frac{\mathrm{Im}\,B_{ii}}{2m_i},
  \label{eq:T}
\end{equation}
and $T_{ji}=-T_{ij}^*$. The amplitude of the $(i,j)$ rotation is
\begin{equation}
  |T_{ij}| \simeq \Theta_{ij}\equiv\frac{\mu\,|N_iN_j|}{|m_j-m_i|},
  \label{eq:dtheta}
\end{equation}
which holds exactly for $m_1=0$ and, more generally, up to corrections of relative order
$|m_j-m_i|/(m_i+m_j)$; it is also the real-symmetric result when the phases vanish.
Equation~\eqref{eq:dtheta} is linear in $\mu\propto1/\Lambda$ and enhanced for nearly
degenerate states. The mixing-angle shifts $\Delta\theta_{ab}^{(1)}$ are obtained by
inserting $U_0(\mathbb{1}+T)$ into the moduli-based definitions of the angles
(Sec.~\ref{sec:numerics}) and expanding to first order. They are linear combinations of
all three amplitudes $T_{12},T_{13},T_{23}$ weighted by entries of $U_0$, so only
$\Delta\theta_{12}$ is well approximated by a single $\Theta_{ij}$ (see
Sec.~\ref{sec:discussion}). For the inputs used below $|N|=(0.952,\,0.687,\,1.273)$. The
expansion breaks down when $\Theta_{ij}$ becomes of order the perturbative limit, that
is, when $\mu|N_iN_j|$ becomes comparable to the relevant mass splitting.

\section{Numerical Results}
\label{sec:numerics}

We use the NuFIT~6.1 normal-ordering best fit including IceCube and Super-Kamiokande
atmospheric data~\cite{nufit61,sharma2026precisionerareassessmentmodifiedquarklepton}: $\sin^2\theta_{12}=0.3088$,
$\sin^2\theta_{13}=0.02248$, $\sin^2\theta_{23}=0.470$, i.e.\
$\theta_{12}=33.76^\circ$, $\theta_{13}=8.62^\circ$, $\theta_{23}=43.28^\circ$, and
$\delta=212^\circ$, together with $\Delta m^2_{21}=7.537\times10^{-5}~\text{eV}^2$ and
$\Delta m^2_{31}=2.511\times10^{-3}~\text{eV}^2$.
The Majorana phases are set to zero. Two spectra are compared. In the quasi-degenerate
(QD) case $m_1 = 0.05$~eV, so $m_2-m_1\simeq7.5\times10^{-4}$~eV and
$m_3-m_1\simeq2.0\times10^{-2}$~eV; this benchmark has $\sum m_\nu\simeq0.17$~eV and is
disfavored by standard cosmological bounds, so it is retained as a limiting illustration
only. In the normal-ordered (NO) case $m_1=0$, so
$m_2-m_1=\sqrt{\Delta m^2_{21}}\simeq8.7\times10^{-3}$~eV and
$m_3=\sqrt{\Delta m^2_{31}}\simeq0.050$~eV ($\sum m_\nu\simeq0.059$~eV). An intermediate
choice $m_1=0.02$~eV gives intermediate shifts (for example
$\Delta\theta_{12}\simeq28^\circ$ at $\Lambda=2\times10^{16}$~GeV).

All results are obtained by exact numerical diagonalization of the $3\times3$ mass
matrix of Eq.~\eqref{eq:corrmass} (Takagi factorization with NumPy/SciPy), not from
first-order perturbation theory. The eigenvectors are matched to the unperturbed mass
eigenstates by maximal overlap, and the corrected angles are read off from the absolute
values of the corrected matrix ($\sin\theta_{13}=|U'_{13}|$,
$\tan\theta_{12}=|U'_{12}/U'_{11}|$, $\tan\theta_{23}=|U'_{23}/U'_{33}|$), which is valid
for any unitary matrix. The corrected matrix is unitary to within $10^{-15}$ and the
Takagi residual is below $10^{-13}$ in every case. The universal coefficient is set to
$c=1$, so absolute shifts scale linearly with $c$.

\begin{figure}[t]
  \centering
  \includegraphics[width=\linewidth]{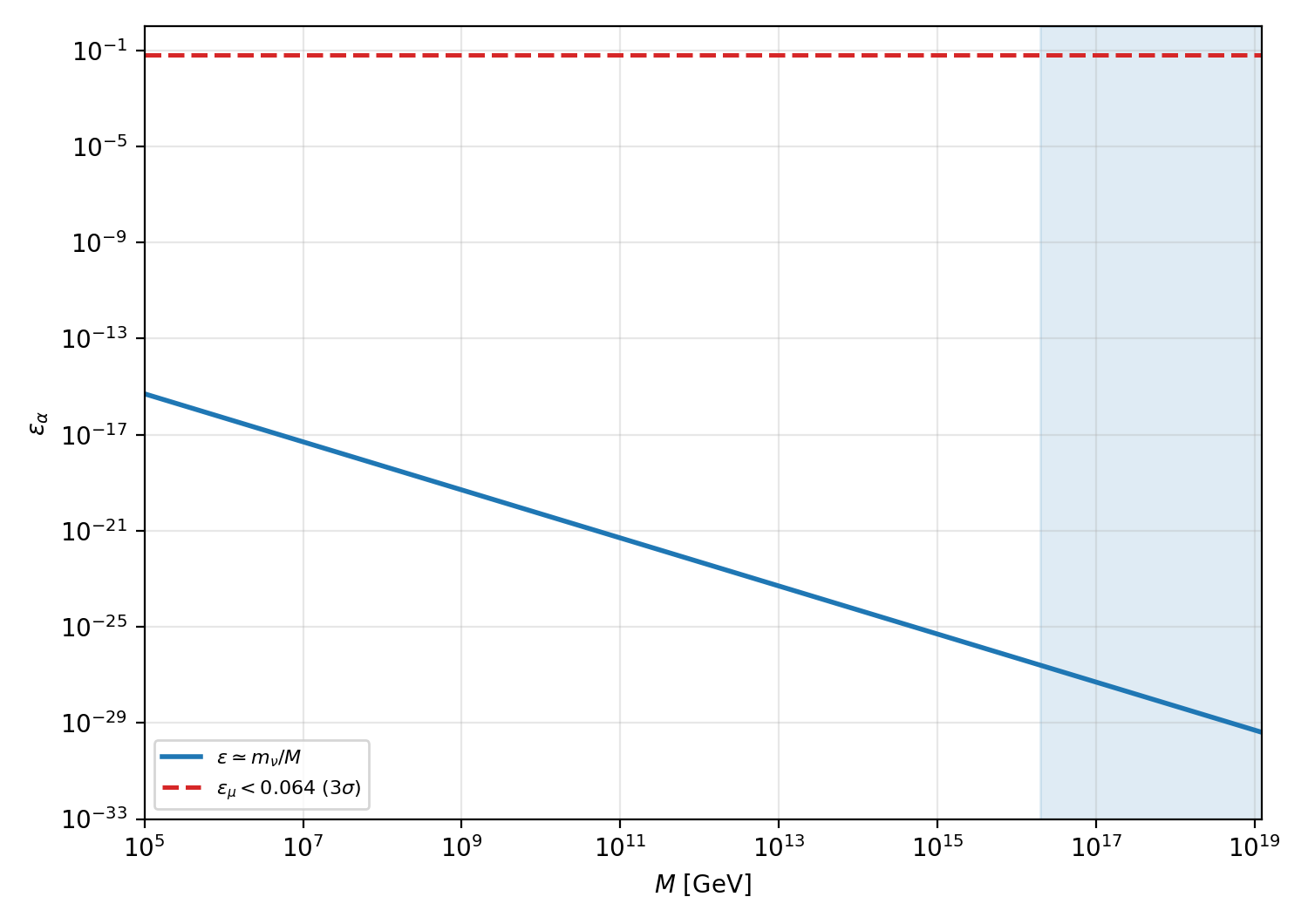}
  \caption{Unitarity violation $\epsilon_\alpha$ of the $3\times3$ block from a
  heavy singlet, Eq.~\eqref{eq:epsheavy} with $m_\nu=0.05$~eV, as a function of
  $M$, against the current $3\sigma$ bound. The shaded band is
  $\MGUT$--$\MPl$.}
  \label{fig:eps}
\end{figure}

\begin{figure}[t]
  \centering
  \includegraphics[width=\linewidth]{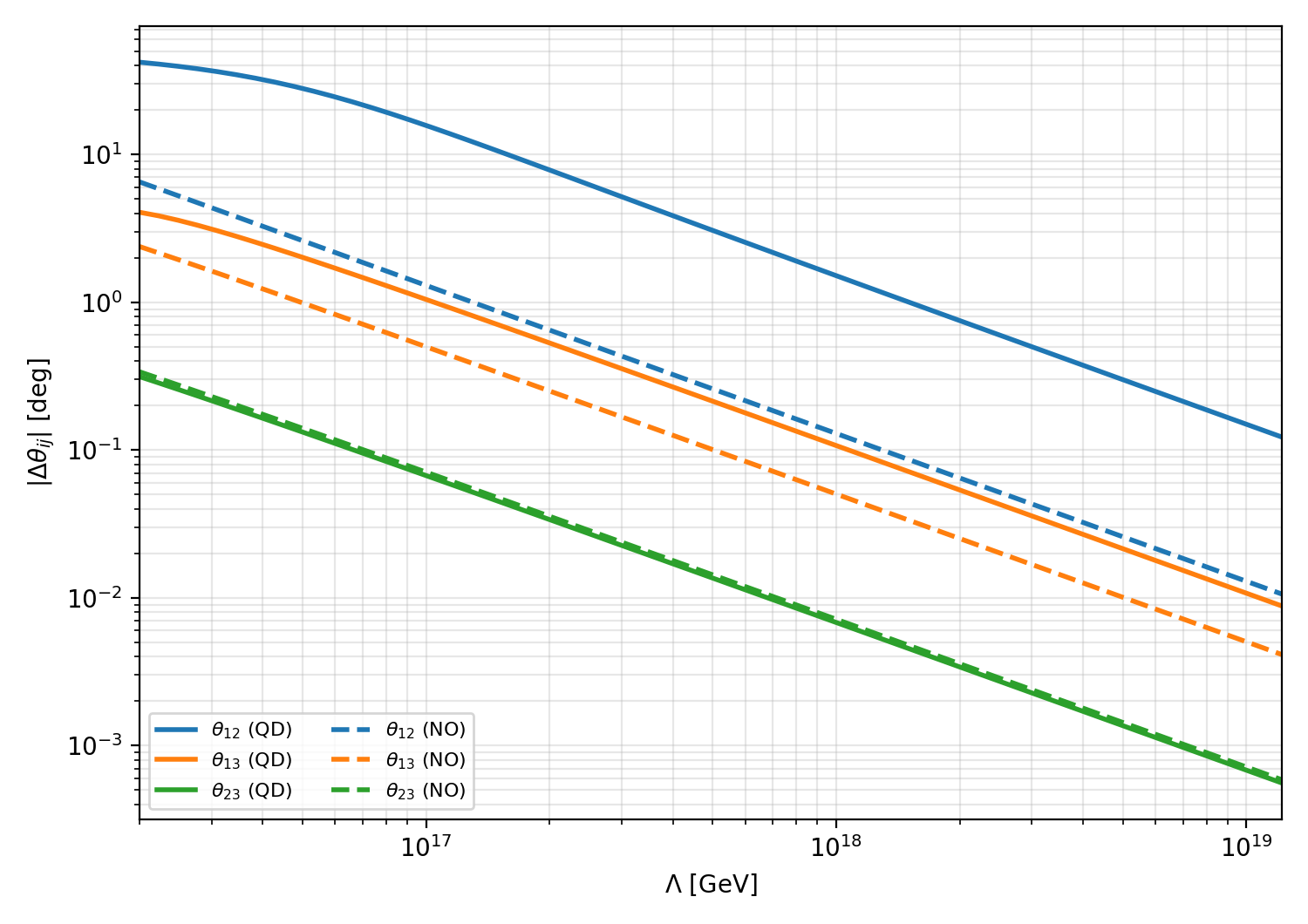}
  \caption{Absolute shifts $|\Delta\theta_{ij}|$ of the three mixing angles as
  functions of $\Lambda\in[\MGUT,\MPl]$ for the QD (solid) and NO (dashed)
  spectra, NuFIT~6.1 inputs. The straight lines on the log-log scale confirm the
  $1/\Lambda$ law away from the nonperturbative region.}
  \label{fig:shifts}
\end{figure}

Several features stand out. (i)~\emph{Spectrum.} For the same $\mu$ the normal ordering
gives much smaller shifts of $\theta_{12}$ than the QD spectrum, because the relevant
splitting $m_2-m_1$ is larger by a factor of about twelve. At $\Lambda=2\times10^{16}$~GeV,
$\Delta\theta_{12}\approx6.5^\circ$ (NO) versus $42^\circ$ (QD). The QD shift is
nonperturbative there, because $\mu\approx1.5\times10^{-3}$~eV exceeds
$m_2-m_1\approx7.5\times10^{-4}$~eV. The NO case is at the edge of the perturbative
regime: although $\mu<m_2-m_1\approx8.7\times10^{-3}$~eV, the enhancement
$|N_1N_2|\simeq0.65$ from the CP phase and the first-octant $\theta_{23}$ raises the
rotation amplitude to $\Theta_{12}\simeq0.11$~rad, and the first-order expansion is no
longer reliable. The expansion parameter is thus $\mu|N_iN_j|/|m_j-m_i|$ rather than
$\mu/|m_j-m_i|$.
(ii)~\emph{Scaling.} In the perturbative regime all shifts follow the $1/\Lambda$ law of
Eq.~\eqref{eq:dtheta} (Fig.~\ref{fig:shifts}). Exact and first-order values agree to
better than $1.5\%$ for $\Delta\theta_{12}$ and $\Delta\theta_{13}$ for the normal ordering
at $\Lambda\gtrsim10^{17}$~GeV and for QD at $\Lambda\gtrsim10^{18}$~GeV, and to better
than $0.2\%$ at $\MPl$. The small shift $\Delta\theta_{23}$, which results from partial
cancellations, is more sensitive to second-order terms: first order is off by about
$8\%$ (NO, $10^{17}$~GeV) and $24\%$ (QD, $10^{18}$~GeV), but within a few percent at
$\MPl$. Table~\ref{tab:theta12} gives the $\theta_{12}$ shifts.
(iii)~\emph{Hierarchy among the angles.} The shifts are ordered
$|\Delta\theta_{12}|>|\Delta\theta_{13}|>|\Delta\theta_{23}|$ for both spectra, and
$\Delta\theta_{13}$ is negative while $\Delta\theta_{12}$ and $\Delta\theta_{23}$ are
positive (Tables~\ref{tab:exact} and~\ref{tab:theta13}); the signs and sizes follow from
the amplitudes $T_{ij}$ and the structure of $U_0$ rather than from any single
$\Theta_{ij}$.
(iv)~\emph{Unitarity.} In every case $\epsilon_\alpha=0$ to numerical precision: the
shifts are pure rotations.

\begin{table}[t]
  \caption{Exact angle shifts at $\Lambda=2\times10^{16}$~GeV
  ($\mu=1.5\times10^{-3}$~eV), NuFIT~6.1 inputs.}
  \label{tab:exact}
  \begin{ruledtabular}
  \begin{tabular}{lcccc}
    Spectrum & $\Delta\theta_{12}$ & $\Delta\theta_{13}$ & $\Delta\theta_{23}$ & Validity \\
    \hline
    QD & $42.1^\circ$ & $-4.07^\circ$ & $+0.31^\circ$ & $\mu>m_2-m_1$ \\
    NO & $6.53^\circ$ & $-2.38^\circ$ & $+0.34^\circ$ & marginal ($\Theta_{12}\approx0.11$) \\
  \end{tabular}
  \end{ruledtabular}
\end{table}

\begin{table}[t]
  \caption{Shifts of $\theta_{12}$ versus $\Lambda$ from exact diagonalization;
  the rotation-amplitude estimates $\Theta_{12}$ of Eq.~\eqref{eq:dtheta} (in degrees)
  are in parentheses.}
  \label{tab:theta12}
  \begin{ruledtabular}
  \begin{tabular}{cccc}
    $\Lambda$ (GeV) & $\mu$ (eV) & QD $\Delta\theta_{12}$ & NO $\Delta\theta_{12}$ \\
    \hline
    $1.22\times10^{19}$ & $2.5\times10^{-6}$ & $0.122^\circ$ ($0.124^\circ$) & $0.0106^\circ$ ($0.0107^\circ$) \\
    $10^{18}$           & $3.0\times10^{-5}$ & $1.51^\circ$ ($1.52^\circ$)   & $0.129^\circ$ ($0.131^\circ$) \\
    $10^{17}$           & $3.0\times10^{-4}$ & $15.7^\circ$ ($15.2^\circ$)   & $1.30^\circ$ ($1.31^\circ$) \\
    $2\times10^{16}$    & $1.5\times10^{-3}$ & $42^\circ$ ($76^\circ$)       & $6.5^\circ$ ($6.5^\circ$) \\
  \end{tabular}
  \end{ruledtabular}
\end{table}

\section{Discussion}
\label{sec:discussion}

\emph{Heavy-state non-unitarity cannot be seen.} Figure~\ref{fig:eps} shows that for
$M\ge\MGUT$ the heavy-state effect lies about 25 orders of magnitude below the bound
(the ratio $0.064/2.5\times10^{-27}\approx2.6\times10^{25}$), so no foreseeable
oscillation experiment can see it. The exception, noted in Sec.~\ref{sec:framework}, is
the case in which $m_D/M$ is tuned independently of $m_\nu$, for example through
cancellations in a multistate seesaw or an approximately conserved lepton number. Then
Eq.~\eqref{eq:epsheavy} fails, the effective non-unitarity can be much larger, and the
bounds of Ref.~\cite{unitarityframework} apply directly to the active--heavy mixing. The
same heavy states also generate a dimension-six operator whose effect on the $3\times3$
block scales as $v^2/\Lambda^2$, far smaller than the dimension-five effect.

\emph{Unitary shifts are larger but invisible to the sum rule.} The shifts of
Table~\ref{tab:exact} are many orders of magnitude larger than $\epsilon$ from
Eq.~\eqref{eq:epsheavy}, but they change angles, not row sums. Since $U'$ is unitary,
$\sum_i|U'_{\alpha i}|^2$ is conserved identically and the sum rule~\eqref{eq:sumrule}
does not feel the shifts at all. They can be probed only through precision measurements
of the angles themselves, compared with predictions of a high-scale model.

\emph{A consistency relation for three flavors.} In a four-flavor framework a high-scale
angle shift is accompanied by a correlated change in the sterile fraction~\cite{koranga2026}.
With three flavors that signature is absent. Its place is taken by a different relation:
at first order every shift is linear in the single parameter $\mu=cv^2/\Lambda$,
$\Delta\theta_{ab}=\mu\,\kappa_{ab}$, where the coefficients $\kappa_{ab}$ are fixed by
the mass spectrum and the unperturbed mixing through Eqs.~\eqref{eq:Ni} and~\eqref{eq:T}.
The ratios $\Delta\theta_{12}:\Delta\theta_{13}:\Delta\theta_{23}$ are therefore
independent of $\Lambda$ and $c$ in the perturbative regime: numerically
$1:-0.39:+0.056$ for the normal ordering and $1:-0.072:+0.0046$ for the QD spectrum at
$\MPl$. These ratios are \emph{not} given by the naive amplitudes $\Theta_{ij}$ of
Eq.~\eqref{eq:dtheta}: although $\Theta_{12}$ reproduces $\Delta\theta_{12}$ to about one
percent, $\Theta_{23}$ overestimates $|\Delta\theta_{23}|$ by a factor of about five
(for the NO spectrum at $2\times10^{16}$~GeV, $1.83^\circ$ versus $0.34^\circ$) and
$\Theta_{13}$ misses the sign and is off by roughly 10--20\% in magnitude, because the
$\theta_{13}$ and $\theta_{23}$ shifts are combinations of all three rotations with
partial cancellations. The first-order pattern computed from Eq.~\eqref{eq:T}, or the
exact values of Tables~\ref{tab:exact} and~\ref{tab:theta13}, is the reliable statement.
For a given mass spectrum, a flavor-blind high-scale origin therefore makes a correlated
prediction for three angle shifts from one free number, which can in principle be tested
once the absolute mass scale and ordering are known; a low-scale origin of an angle
anomaly generally would not respect it.

\emph{Experimental reach.} The largest shift at $\MPl$, $\Delta\theta_{12}\simeq0.12^\circ$
(QD; $0.011^\circ$ for NO), is well below the present $1\sigma$ precision on $\theta_{12}$
(about $0.7^\circ$~\cite{esteban2024}), so Planck-scale effects of this type are out of
reach in the angles as well. For the QD spectrum and $\Lambda\lesssim10^{17}$~GeV the
shifts reach several to tens of degrees and would far exceed the sensitivity of precision
experiments. For the normal ordering the $\theta_{12}$ shift becomes comparable to or
larger than the present uncertainty for $\Lambda\lesssim10^{17}$~GeV ($1.3^\circ$ at
$10^{17}$~GeV) and reaches several degrees for $\Lambda$ close to $\MGUT$
(Table~\ref{tab:exact}). Both statements depend on $c$ and on the absolute neutrino mass
scale, neither of which is known.

\emph{Renormalization-group running.} Renormalization-group (RG) running between
$\MGUT$ and the electroweak scale modifies the mixing angles through continuous radiative
corrections, which are known to be larger for quasi-degenerate spectra than for
hierarchical ones~\cite{chankowski1993,babu1993,antusch2003}. In contrast, higher-dimension
operators at $\Lambda$ generate threshold shifts proportional to $v^2/\Lambda$. RG running
is not included in the numerics here and would have to be combined with the operator
shifts in any quantitative comparison to a specific high-scale model.

\begin{table}[t]
  \caption{Exact shifts of $\theta_{13}$ and $\theta_{23}$ (degrees) for the QD and
  normal-ordered spectra.}
  \label{tab:theta13}
  \begin{ruledtabular}
  \begin{tabular}{ccccc}
    & \multicolumn{2}{c}{QD} & \multicolumn{2}{c}{NO} \\
    $\Lambda$ (GeV) & $\Delta\theta_{13}$ & $\Delta\theta_{23}$ & $\Delta\theta_{13}$ & $\Delta\theta_{23}$ \\
    \hline
    $1.22\times10^{19}$ & $-0.0088$ & $+0.00056$ & $-0.0041$ & $+0.00059$ \\
    $10^{18}$           & $-0.107$  & $+0.0068$  & $-0.050$  & $+0.0072$ \\
    $10^{17}$           & $-1.04$   & $+0.067$   & $-0.50$   & $+0.071$ \\
    $2\times10^{16}$    & $-4.07$   & $+0.31$    & $-2.38$   & $+0.34$ \\
  \end{tabular}
  \end{ruledtabular}
\end{table}

\emph{Caveats.} The numerics use a flavor-democratic all-ones matrix, the NuFIT~6.1
central values, and a single Dirac phase; the Majorana phases are set to zero. Because
$N_i=\sum_\alpha U_{0,\alpha i}$ depends on the column phases of $U_0$, the shifts depend
on the Majorana phases and on the sign and phase of the coefficient $c$ relative to the
mass matrix, not only on $\delta$; the results here should be read as a benchmark, not as
a bound over the phase space. The $\theta_{12}$ shift in particular is sensitive to the
phase-dependent combination $|N_1N_2|$, which for the NuFIT~6.1 inputs ($|N_1N_2|=0.65$)
is nearly three times larger than for the earlier illustrative choice $\delta=0$,
$\theta_{23}=49^\circ$ ($0.24$). The universal coefficient $c$ is set to one, so the
$1/\Lambda$ scaling is robust while the absolute values are not. Mixing angles and
masses are best-fit values without uncertainties, which are not propagated. A
non-universal $c_{\alpha\beta}$ would change the vector $N$ in Eq.~\eqref{eq:Ni} and hence
the ratios discussed above. Finally, with $c=1$ and $\Lambda=\MGUT$ the operator has
$\mu\simeq1.5$~meV, which is comparable to the light masses themselves; treating it as a
small correction assumes that the dominant neutrino mass arises from a different origin.

\section{Conclusion}
\label{sec:conclusion}

Above $\MGUT$ the unitarity of the $3\times3$ PMNS matrix is violated at most at the
level $m_\nu/M\sim10^{-27}$, and a sum-rule test is blind to the dominant high-scale
effect, a set of exactly unitary angle shifts of size up to a few tenths of a degree
at $\MPl$ ($0.12^\circ$ for QD, $0.011^\circ$ for NO, with NuFIT~6.1 inputs) and growing
as $1/\Lambda$ toward $\MGUT$. A direct test of three-flavor
unitarity therefore constrains only low-scale extensions of the lepton sector, while the
scale above $\MGUT$ can be accessed only through precise angle measurements. Heavy-state
non-unitarity and Planck-type unitary shifts are different phenomena with different
observables and should not be conflated; the latter scale as $v^2/\Lambda$, become
nonperturbative when $\mu|N_iN_j|$ approaches the relevant mass splitting, and in the
three-flavor case are tied together by a single parameter through the fixed
$\Lambda$-independent ratios of the angle shifts.

Natural extensions include scanning the Dirac and Majorana phases and non-universal
coefficients, including RG running in the numerics, and propagating present angle
uncertainties to estimate the discovery reach for a given $\Lambda$.

\bibliographystyle{apsrev4-2}
\bibliography{references}

\end{document}